# In-line fiber optic optofluidic sensor based on a fully open Fabry-Perot interferometer

Dewen Duan[1,*], Qian Kang[1], Qianhui Yang[1], Zihao Zhao[1], Na Li[2], Guan-Xiang Du[2], Yi-Yuan Xie[1,*]

[1] School of Electronic and Information Engineering, Southwest University, No.2, Tiansheng Road, BeiBei District, Chongqing 400715, China
[2] School of Communication and Information Engineering, Nanjing University of Posts and Telecommunications, No. 66 Xin Mofan Road, Nanjing 210000, China

E-mail: ddw225@swu.edu.cn  and  yiyuanxie@swu.edu.cn

## Abstract

We present an all-fiber, fully open Fabry-Perot interferometer (FPI) cavity that is suitable for fluidic measurement applications. Fabrication of the FPI involves the alignment and bonding of three optical fiber sections using either ceramic glue or low-temperature melting glass. The fabrication procedure allows the protection of the cleaved optical fiber end faces, which serve as the two mirrors of the FPI, from damage, thus ensuring the high visibility of the FPI sensor. The FPI's complete openness permits the analyte of interest fluids to flow smoothly into the cavity and interact directly with the light, obviating the need for additional assistance. The fabrication experiment demonstrates that the fabrication procedure can readily achieve a visibility of over 20 dB. Refractive index testing indicates that the sensor exhibits a sensitivity of over 1116 nm/RIU within the range of 1.334-1.375. A comparison of temperature investigations indicates that the fully open cavity FPI fabricated by bonding with low-temperature melting glass exhibits relatively lower temperature immunity than that bonded with ceramic glue. Both have a relatively low temperature fluctuation within the temperature range of 40°C-100°C, with less than 3 nm and 4.5 nm in the over 60°C changes, respectively. Our proposed fully open FPI is an economical, robust, and simple-to-fabricate structure with the potential for mass production. This renders it an appealing option for practical optofluidics applications.

## 1. Introduction

The optical fiber Fabry-Perot interferometer (FPI) has been investigated extensively due to its numerous advantages. These include high-precision measurement capabilities, immunity to electromagnetic interference, and the potential for mass multiplexing[1]. The optical fiber Fabry-Perot interferometer (FPI) has been the subject of extensive investigation in a number of fields, including temperature measurement[2-4], strain or stress monitoring[5-7], pressure detection[8], displacement intorgation[9], refractive index investigation[10-13], and others. To achieve these specialized application goals, many optical fiber FPI configurations have been proposed. For instance, to measure temperature, an FPI with a cavity filled with polymers is proposed for low-temperature range measurements[3], while a sapphire fiber cavity is proposed for high-temperature measurements[4]. To monitor stress, an in-line FPI is proposed, formed by splicing a hollow core fiber between two sections of single-mode fiber[5,7]. To integrate displacements, an FPI with a moveable mirror formed by a sphere ball is proposed[9]. In particular, lots of open cavity FPIs have been employed to investigate the fluidic properties.

The most straightforward open-cavity FPI is directly machined on optical fiber using modern microfabrication techniques. For instance, femtosecond lasers have been utilized to fabricate open-cavity FPI directly on optical fibers[6,10,13]. Similarly, focused ion beams have been employed to fabricate open-cavity FPI directly on fibers[14,15]. In addition, diamond blades have been used to dice open-cavity FPI directly on fibers[16,17]. Another type of open cavity FPI requires a somewhat additional procedure to fabricate. First, a bubble is formed inside optical fibers by splicing. Then, a channel is drilled to the bubble to form a gas or fluid access channel by laser machining[18]. The same procedure has also been employed in mechanical polishing or laser ablation of a gas or fluid access channel through FPI, formed by splicing short sections of hollow-core fibers between solid-core fibers[12,19]. In addition to micromachining, splicing has also been employed in the fabrication of an open-cavity FPI. For instance, a C-shaped optical fiber has been spliced between single-mode fibers to form an open cavity FPI[20,21]. Lateral offset splicing has also been investigated as a method for splicing open cavity FPI[11,22]. Offset splicing involves the splicing of a small fiber between a single-mode fiber or between a single-mode fiber and a seven-core fiber to form an open cavity FPI[23,24]. There are also many demonstrations of subsequent splicing of hollow core fibers/capillaries and small hole fibers or microstructured fibers to form open cavity FPIs [25-27].

Here we propose an all-fiber, fully open FPI suitable for fluidic applications. Fabrication of the FPIs involves the alignment and bonding of three sections of optical fiber with either ceramic glue or low-temperature melting glass. The fabrication procedure ensures the protection of cleaved optical fiber end faces from damage, thereby facilitating the observation of the FPI sensor's high visibility of the reflection spectrum. Moreover, the FPI is entirely open, enabling the unimpeded flow of analyte-of-interest fluids into the cavity and direct interaction with the light, thus eliminating the necessity for additional assistance.

## 2. Sensor principles and fabrication

*2.1 Sensor working principles.* The principle schematic of the fully open FPI is shown in Fig.1. The two reflection mirrors of the FPI are formed by the reflection fiber end face and the lead-in fiber end face, respectively. The gap between the two end faces formed the cavity of the fullyd open FPI. The reflectivity of the optical fiber end face in analyte fluid is

$$R = \left(\frac{n_o - n_f}{n_o + n_f}\right)^2 \quad (1)$$

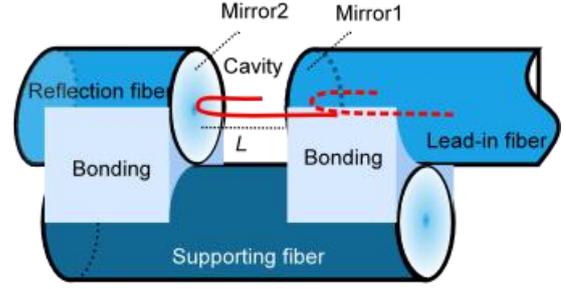

fig. 1. The principle schematic of the fully open Fabry-Perot interferometer (FPI).

$n_o$ = 1.45 is the refractive index of the optical fiber core, and $n_f$ is the refractive index of the analyte fluid inside the FPI cavity. Then, the reflected total light field of the two mirrors of the FPI is

$$E_r = E_0\left(\sqrt{R} + (1-\alpha)(1-R)\sqrt{R}e^{-j(2\phi+\varphi_0)}\right) \quad (2)$$

$\alpha$ is the ratio of transmission loss caused by the light beam divergence in the cavity and the lead-in optical fiber end surface imperfection, $\phi$ is the round-trip propagation phase shift given by $\phi = (2\pi n_f L)/\lambda$, $L$ is the length of the FPI cavity and $\varphi_0$ is the initial phase. When

$$2\phi + \varphi_0 = (4\pi n_f L)/\lambda_v + \varphi_0 = 2\pi(m + 1/2) \quad (3)$$

$E_r$ reached its minimum value and the reflective spectrum shows its valley. Where $m$ is a natural number with a value of $m$ = 0, 1, 2, 3, ... Taking the derivative of $n$ with respect to $\lambda_v$, we obtain

$$\Delta\lambda_v/\Delta n_f = 4\pi L \times \text{constant} \quad (4)$$

Assume that the FPI cavity $L$ is constant. From Eq. 4 we get

$$\Delta\lambda_v/\lambda_v = \Delta n_f/n_f \quad (5)$$

Eq.5 shows that we can use the FPI's interference valley $\lambda_v$ to monitor/calculate the refractive index change of the analyte fluid.

*2.2 Sensor fabrication.* The sensor is fabricated by first cutting three sections of single-mode optical fiber. One is a longer section that serves as the input fiber. The remaining two sections are short and serve as the support and reflection fibers, respectively. The length of the support fiber is approximately 2 to 4 millimeters, while the length of the reflection fiber is approximately 1 millimeter. Secondly, the three sections of fibers are aligned on a glass plate, with the reflection fiber and the lead-in fiber aligned alongside the support fiber with a gap between their end faces, forming the FPI cavity. The reflection fiber and the lead-in fiber each touch approximately half the length of the support fiber. As illustrated in Fig.2(a), the third step involves the application of a small quantity of ceramic glue or glass slurry (Low melting point glass powder mixed with water, the melting point of the glass powder is 450°C-550°C.) to the contact line of the supporting fiber and the lead-in fiber, and the contact line of the supporting fiber and the reflection fiber, respectively (Fig.2(b)). Subsequently, the ceramic glue is permitted to solidify, or the low-melting glass slurry is heated to melt and then cooled (Fig.2(c)). The microscopic images of two fabricated FPI are shown in Fig.3. Fig.4 shows the reflection spectrum of the glass melt-bonded FPI (with a cavity length of approximately 74 μm and ceramic glue-bonded FPI (with a cavity length of approximately 71 μm in air and water respectively.

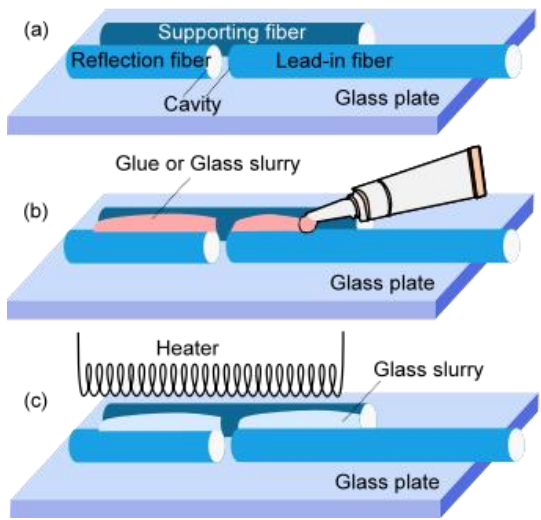

Fig. 2. The fabrication processing of the fully open FPI. (a) Three sections of fibers are aligned on a glass plate. (b) Apply ceramic glue or glass slurry to the contact line of the support fiber and the lead-in fiber. (c) Heating of low-melting glass slurry to melting point.

## 3. Experimental results and discussion

The open cavity FPI sensor's performance investigation setup is depicted in Fig.5. The broadband light from the white light source was launched into the open cavity FPI sensor through the optical fiber connected to the optical circulator. The reflection spectrum of the FPI was recoupled to the optical circulator and recorded using an optical spectrum analyzer (OSA, Anritsu, MS9740A). The open cavity FPI sensor was immersed in the analyte fluid in a beaker on the thermostatic heating table.

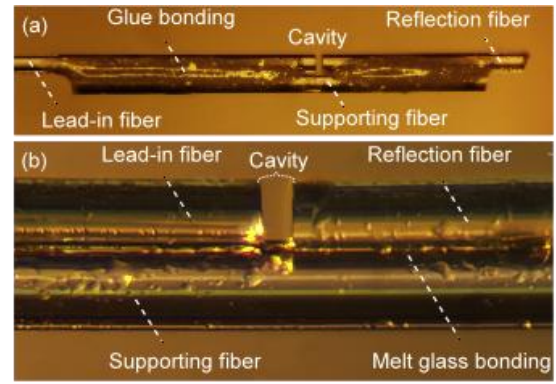

Fig. 3. The photos of fabricated fully open FPI (a) bonded with ceramic glue, and (b) bonded with low melt point glass.

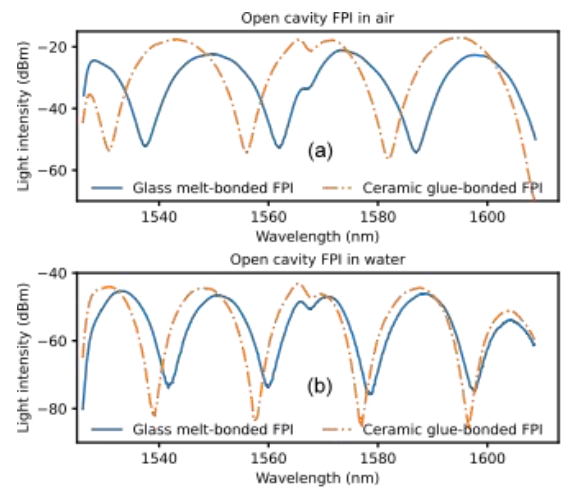

Fig. 4. The reflection spectrum of fabricated fully open FPIs (illustrated in Fig.3) when they are immersed in (a) air and (b) water. The cavity length of the glass melt-bonded FPI is approximately 74 μm, while the cavity length of the ceramic glue-bonded FPI is approximately 71 μm.

Initially, the open cavity FPI sensors' refractive index performance was investigated by measuring a group of glucose solutions with refractive index values ranging from 1.3346 to 1.3745. The refractive index values of the glucose solutions were calibrated using an Abbe refractometer (Shanghai Li Chen, accuracy: 0.0002 nD). This investigation was conducted at room temperature without the necessity of heating the beaker. The reflection spectra of the glass melt-bonded FPI (with a cavity length of approximately 74 μm) and ceramic glue-bonded FPI (with a cavity length of approximately 71 μm) immersed in glucose solutions with different refractive indices are presented in Fig.6. The shifts of the reflective spectrum valleys of the two open cavity FPIs corresponding to the refractive index changes are illustrated in Fig.7. This figure demonstrates that valleys at a longer wavelength have a higher refractive index measuring sensitivity for the same open cavity FPI. This is consistent with Eq.5. The highest sensitivity is 1160.4 nm/RIU (refractive index unit), while the lowest sensitivity is 1116.5 nm/RIU.

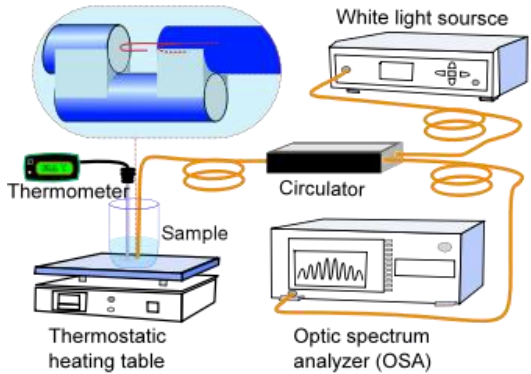

Fig. 5. The open cavity FPI sensor performance investigation setup.

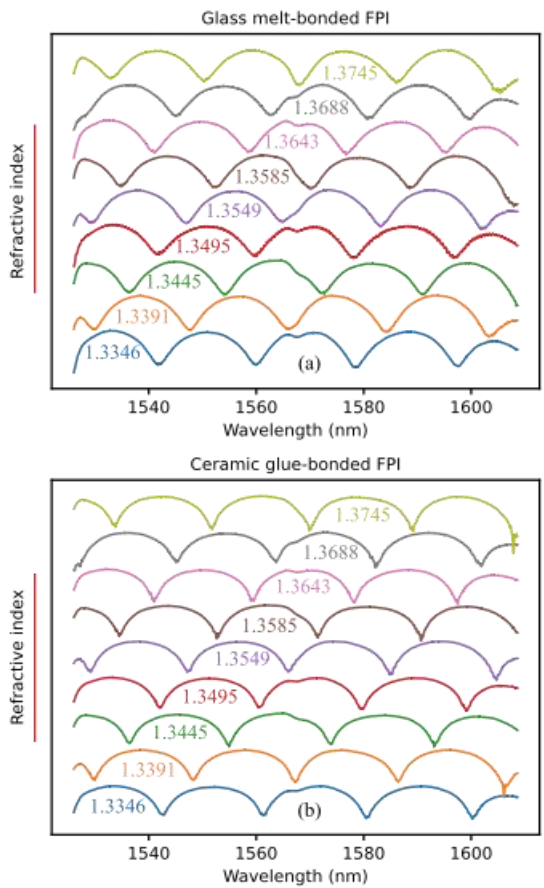

Fig. 6. The reflection spectra of the glass melt-bonded FPI and the ceramic glue-bonded FPI when immersed in glucose solutions with varying refractive indices.

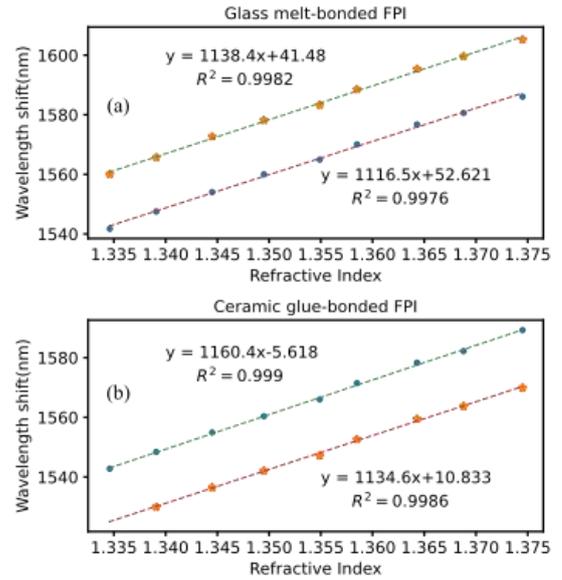

Fig. 7. The reflection spectrum valleys of (a) the glass melt-bonded FPI and (b) the ceramic glue-bonded FPI exhibit shift corresponding to refractive index changes when immersed in glucose solutions.

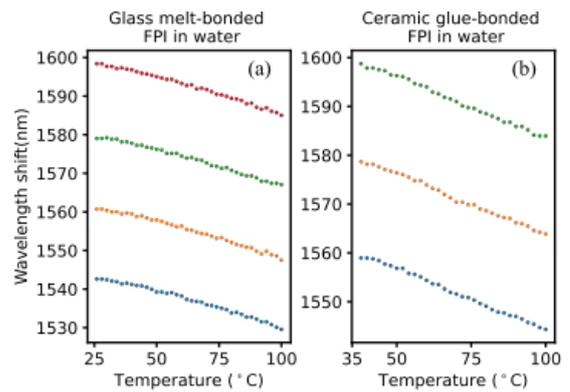

Figure 8: The reflection spectrum valleys of (a) the glass melt-bonded FPI and (b) the ceramic glue-bonded FPI immersed in water exhibit shift corresponding to the temperature changes of the water when immersed in water.

Additionally, the temperature-dependent refractive index of tap water was measured using the experimental setup depicted in Fig.5. The open cavity FPIs were immersed in the tap water inside the beaker. The tap water inside the beaker was heated to a temperature of 100°C, as indicated by the thermometer. Subsequently, the heater was deactivated in order to permit the water to cool gradually while the interference spectrum was recorded at 2-degree intervals as the temperature decreased to ambient temperature. Fig.8 depicts the reflective spectrum valleys of the two open cavity FPIs as a function of the water's temperature. The temperature increase results in a shift of the interference fringe to a shorter wavelength, which can be interpreted as a decrease in the refractive index. This is consistent with the findings of previous studies [14].

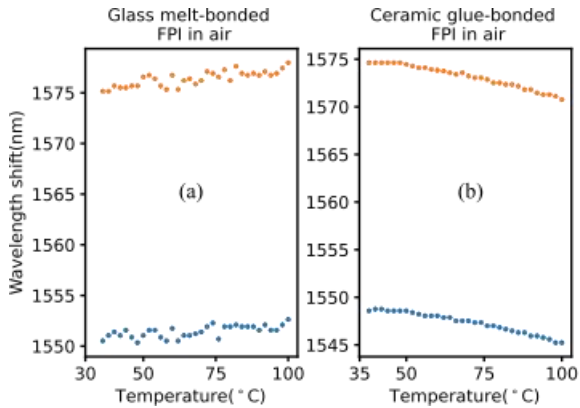

Fig. 9. The reflection spectrum valleys of (a) the glass melt-bonded FPI and (b) the ceramic glue-bonded FPI in glass tubes exhibit shifts corresponding to the temperature changes.

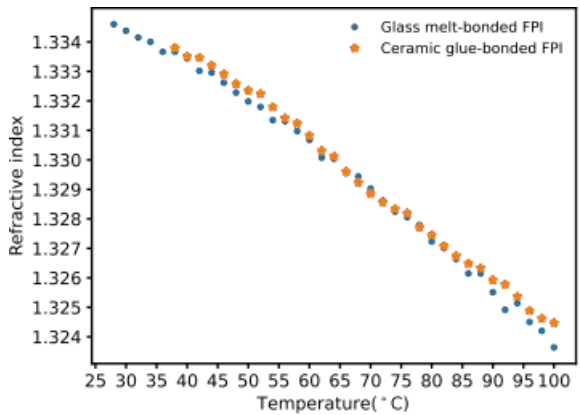

Fig. 10. The calculated refractive index of tap water at varying temperatures. The results were calculated using Eq. 5 and the data presented in Fig.8.

Later, we investigated the temperature response of the two fabricated open cavity FPI by put the FPIs inside a one-end sealed glass tube with inner and outer diameters of approximately 300 μm and 500 μm respectively to protect it from contact with the water. The tube was then immersed in the water bath inside the beaker on the thermostatic heating table. We repeated the first heating of the water inside the beaker to 100°C and then turned off the heater procedure. We recorded the interference spectrum at 2-degree intervals as the temperature decreased from 100°C to the ambient temperature. Fig.9 illustrates the reflective spectrum valleys of the two open cavity FPIs as a function of temperature. The results demonstrate that from room temperature to 100°C, the valleys of the glass melt-bonded FPI exhibit a maximum shift of 2.816nm, while the valleys of the ceramic glue-bonded FPI exhibit a maximum shift of 4.224nm. In contrast, for the glass melt-bonded FPI, the spectral valleys shift to longer wavelengths with increasing, while for the ceramic glue-bonded FPI, the spectral valleys shift to shorter wavelengths with increasing, which is similar to our previous finding in femtosecond machined FPI in optical fiber[6]. It is hypothesized that the shift in the spectral valleys of glass melt-bonded FPI is caused by the thermal expansion of the supporting optical fiber, while the shift in the ceramic glue-bonded FPI is caused by the combined effect of the thermal expansion of the ceramic glue-bonded, shortens the cavity length, and the thermal expansion of the supporting optical fiber, which elongates the cavity length.

The refractive index of tap water at different temperatures can be calculated using Eq. 5 and the data presented in Fig.8. The results are shown in Fig. 10. The results demonstrate that the tendency and amount of change are consistent with the findings of previous studies of ionized water[10]. The data exhibits a certain degree of irregularity, likely due to the precision of our thermometer, which is approximately 2°C. This irregularity is also reflected in Fig.8. Nevertheless, the experiment demonstrates that our presented sensor can be utilized in optofluidics applications.

## 4. Conclusion

We demonstrated a fully open Fabry-Perot interferometer (FPI) cavity suitable for fluidic measurement applications. The FPI is fabricated by aligning and bonding three sections of optical fiber using either ceramic glue or low-temperature melting glass. The fabrication procedure ensures the protection of the cleaved optical fiber end faces, which serve as the two mirrors of the FPI, from damage. This allows for high visibility of the FPI sensor. The complete openness of the FPI allows the analyte of interest to easily flow in and out of the cavity and interact directly with the light, eliminating the need for additional assistance. Two open cavity FPIs with a cavity length of approximately 70 μm were fabricated experimentally, one bonded by glass melt bond and the other by ceramic glue bond. Both the glass-bonded and ceramic-bonded FPIs exhibited a visibility of over 25 dBm. Furthermore, the refractive index performance of the FPIs was investigated through experimentation. The results demonstrated that the sensors exhibited a sensitivity of over 1116 nm/RIU in the refractive index range of 1.334-1.375. A comparison of temperature investigations indicated that the open cavity FPI fabricated by bonding with low-temperature melting glass exhibited relatively lower temperature immunity than that bonded with ceramic glue. Both FPIs exhibited minimal temperature-induced shifts in their spectral valleys, ranging from 2.816 nm to 4.224 nm, across a temperature range of 40°C to 100°C. In the case of the glass melt-bonded FPI, the spectral valleys shift to longer wavelengths as the temperature increases, whereas in the ceramic glue-bonded FPI, the spectral valleys shift to shorter wavelengths as the temperature increases. The proposed fully open FPI is an economical, robust, and simple-to-fabricate structure with the potential for mass production. This renders it an appealing option for practical optofluidics applications.

## Conflict of interest

The authors declare that they have no known competing financial interests or personal relationships that could have appeared to influence the work reported in this paper

## Funding

This work is supported by the Chongqing Ph.D. "Through Train" Scientific Research Project (Grant Number: sl202100000172) and Fundamental Research Funds for the Central Universities (Grant Number: SWU021004).